\documentclass[aps,prl,floatfix,superscriptaddress,reprint]{revtex4-2}
\usepackage[utf8]{inputenc}
\usepackage[T1]{fontenc}
\usepackage{amssymb}
\usepackage{bm}
\usepackage{graphicx}
\usepackage{textgreek}
\usepackage[separate-uncertainty=true]{siunitx}
\usepackage[version=4]{mhchem}
\usepackage[hidelinks]{hyperref}
\usepackage[capitalize,nameinlink]{cleveref}

\DeclareSIQualifier\rms{RMS}
\DeclareSIUnit\currdens{\ampere\per\square\micro\metre}

\begin{document}
\title{
    Disentangling electrical switching of antiferromagnetic NiO using high magnetic fields
    % Suppressing electrical switching of antiferromagnetic NiO using high magnetic fields
}
\date{\today}

\newcommand{\TUe}{Department of Applied Physics, Eindhoven University of Technology, P.O.\ Box 513, 5600 MB Eindhoven, the Netherlands}
\newcommand{\HFML}{High Field Magnet Laboratory (HFML - EMFL), Radboud University, 6525 ED Nijmegen, The Netherlands}
\newcommand{\ITP}{Institute for Theoretical Physics, Utrecht University, Leuvenlaan 4, 3584 CE Utrecht, The Netherlands}

\newcommand{\IEP}{Institute of Experimental Physics, Faculty of Physics, University of Warsaw, ul.\ Pasteura 5, PL-02-093 Warsaw, Poland}
\newcommand{\MagLab}{National High Magnetic Field Laboratory, Los Alamos National Laboratory, Los Alamos, NM 87545, USA}

\newcommand{\txtclrA}{green}
\newcommand{\txtclrB}{orange}

\author{C.\,F.\ Schippers}
\email{c.f.schippers@tue.nl}
\affiliation{\TUe}

\author{M.\,J.\ Grzybowski}
\altaffiliation[Current address: ]{\IEP}
\affiliation{\TUe}

\author{K.\ Rubi}
\altaffiliation[Current address: ]{\MagLab}
\affiliation{\HFML}

\author{M.\,E.\ Bal}
\affiliation{\HFML}

\author{T.\,J.\ Kools}
\affiliation{\TUe}

\author{R.\,A.\ Duine}
\affiliation{\TUe}
\affiliation{\ITP}

\author{U.\ Zeitler}
\affiliation{\HFML}

\author{H.\,J.\,M.\ Swagten}
\affiliation{\TUe}

\begin{abstract}
    % Version 1
    % Recent demonstrations of the electrical switching of thin-film antiferromagnets (AFs) have given an enormous impulse to the field of AF spintronics.
    % However, many of these observations are plagued by non-magnetic, parasitic effects, caused by the high current densities needed for the experiments; these non-magnetic effects can lead to very similar behavior and are very difficult to distinguish from the actual magnetic effects.
    % Often, resolving these two effects requires techniques to image the antiferromagnetic domains after switching.

    % In this investigation on \ce{Pt/NiO} devices, we use magnetic fields up to \SI{15}{\tesla} as an alternative way to quantitatively disentangle the magnetic and non-magnetic effects in experiments aiming at switching the \SI{5}{\nano\metre} thick antiferromagnetic \ce{NiO} layer.
    % We demonstrate that these high magnetic fields of over \SI{13.5}{\tesla} suppress the electrical switching of \ce{NiO}, but leave the non-magnetic contribution to the signal intact.
    % Using a monodomainization model the magnetic and non-magnetic contributions are separated from each other, showing how they behave as a function of the current density.

    % These results demonstrate that combining electrical methods and strong magnetic fields can be an invaluable tool for antiferromagnetic spintronics.
    % It allows for implementing and studying electrical switching of antiferromagnets in more complex systems where the antiferromagnet is not accessible for imaging.

    % Version 2
    Recent demonstrations of the electrical switching of antiferromagnets (AFs) have given an enormous impulse to the field of AF spintronics.
    Many of these observations are plagued by non-magnetic effects that are very difficult to distinguish from the actual magnetic ones.
    Here, we study the electrical switching of thin (\SI{5}{\nano\metre}) \ce{NiO} films in \ce{Pt/NiO} devices using magnetic fields up to \SI{15}{\tesla} to quantitatively disentangle these magnetic and non-magnetic effects.
    We demonstrate that these fields suppress the magnetic components of the electrical switching of \ce{NiO}, but leave the non-magnetic components intact.
    Using a monodomainization model the contributions are separated, showing how they behave as a function of the current density.
    These results show that combining electrical methods and strong magnetic fields can be an invaluable tool for AF spintronics, allowing for implementing and studying electrical switching of AFs in more complex systems.
\end{abstract}

\maketitle

%   _____   _   _   _______   _____     ____    _____    _    _    _____   _______   _____    ____    _   _
%  |_   _| | \ | | |__   __| |  __ \   / __ \  |  __ \  | |  | |  / ____| |__   __| |_   _|  / __ \  | \ | |
%    | |   |  \| |    | |    | |__) | | |  | | | |  | | | |  | | | |         | |      | |   | |  | | |  \| |
%    | |   | . ` |    | |    |  _  /  | |  | | | |  | | | |  | | | |         | |      | |   | |  | | | . ` |
%   _| |_  | |\  |    | |    | | \ \  | |__| | | |__| | | |__| | | |____     | |     _| |_  | |__| | | |\  |
%  |_____| |_| \_|    |_|    |_|  \_\  \____/  |_____/   \____/   \_____|    |_|    |_____|  \____/  |_| \_|

% \section{Introduction}
Recently, there has been a lot of attention for controlling the magnetic order of antiferromagnetic materials.
The insensitivity to external magnetic fields, combined with \si{\tera\hertz}-frequency magnetization-dynamics, makes antiferromagnets interesting for numerous applications, ranging from data-storage devices~\cite{Jungwirth2016} to \si{\tera\hertz} radiation sources~\cite{Khymyn2017,Stremoukhov2019,Cheng2016}.  % Cheng??
However, control over the orientation of the magnetic order in an antiferromagnet remains problematic, and it has only recently been demonstrated that the magnetic order can be controlled using electrical currents~\cite{Wadley2016,Chen2018b}.
Since these first demonstrations, there have been many experiments that manipulate the antiferromagnetic state in a variety of systems, such as multi-layer systems with antiferromagnets and ferromagnets~\cite{Li2020,Liu2020}, or systems with non-collinear antiferromagnets~\cite{Tsai2020,Deng2021}.

In insulating antiferromagnets, such as \ce{NiO}, \ce{CoO}, and \ce{Fe2O3}, experiments have shown that the magnetic order can be controlled using electrical current pulses through an adjacent heavy metal (e.g.\ \ce{Pt})~\cite{Chen2018b,Grzybowski2022,Cheng2020,Moriyama2018}.
The current in the heavy metal layer is, via the spin-Hall effect, converted into a transverse spin current.
In turn, this spin current is injected into the antiferromagnet where it exerts a spin-torque on the spins, which is then expected to manipulate the magnetic order of the antiferromagnet~\cite{Chen2018b,Baldrati2019}.
Alternatively, Joule heating due to the current pulse can give rise to a thermomagnetoelastic effect, which can change the anisotropy of the antiferromagnet sufficiently to induce a change in the antiferromagnetic order~\cite{Meer2021}.

However, recently there has been a debate about the actual origin of the signals observed in the above-mentioned electrical switching experiments, as they can equally well be explained by non-magnetic, parasitic effects, such as structural changes or damage caused by Joule heating, or electromigration~\cite{Chiang2019,Churikova2020,Matalla-Wagner2020,Cheng2020}.
Although there are demonstrations that show with certainty that actual magnetic reorientation is possible, it is generally difficult to distinguish between the magnetic and non-magnetic contributions to the electrical switching signals, as the electrical measurements themselves consist only of Hall measurements showing the electrically-induced switching as an up-down pattern.
This shortcoming can be resolved using imaging techniques that can resolve the magnetic state of an antiferromagnet, such as XMLD-PEEM~\cite{Baldrati2019,Wadley2018}.
However, this requires an X-ray beam-line and is limited to devices with relatively free access to the antiferromagnet in question, i.e.\ devices where the antiferromagnetic layer is not buried beneath other layers.

Here we disentangle magnetic and non-magnetic effects in the electrical switching of thin films of \ce{NiO} using strong magnetic fields of up to \SI{15}{\tesla}.
If the magnetic field is stronger than the monodomainization field (determined to be \SI{13.5(2)}{\tesla} for our samples, Supplemental Material~\cite{Supplementary}) the magnetic order of \ce{NiO} is organized in a single domain, whose magnetic orientation is controlled by the external magnetic field~\cite{Fischer2018}.
Hence, the effect of reorienting the magnetic moments in an electrical switching experiment is expected to be suppressed in such a field.
As Joule heating and electromigration are not affected by the magnetic field, they will therefore remain present, giving a way to disentangle the magnetic and non-magnetic contributions.

We show that it is indeed possible to suppress the electrical switching effects in \ce{Pt/NiO} devices by applying a sufficiently strong magnetic field of over \SI{13.5}{\tesla}.
Field dependence of the switching signal is observed and can be understood and modelled with a multi-domain interpretation of the NiO magnetic structure~\cite{Fischer2018}.
Using this model, we demonstrate that the magnetic and non-magnetic contributions to the observed switching signal can be separated from each other.
Thereby, this technique helps to better understand the electrical switching experiments and allows integrating and investigating the electrical switching of antiferromagnets in more complex devices.

%   __  __   ______   _______   _    _    ____    _____     _____
%  |  \/  | |  ____| |__   __| | |  | |  / __ \  |  __ \   / ____|
%  | \  / | | |__       | |    | |__| | | |  | | | |  | | | (___
%  | |\/| | |  __|      | |    |  __  | | |  | | | |  | |  \___ \
%  | |  | | | |____     | |    | |  | | | |__| | | |__| |  ____) |
%  |_|  |_| |______|    |_|    |_|  |_|  \____/  |_____/  |_____/

% \section{Methods}
\begin{figure}
    \centering
    \includegraphics{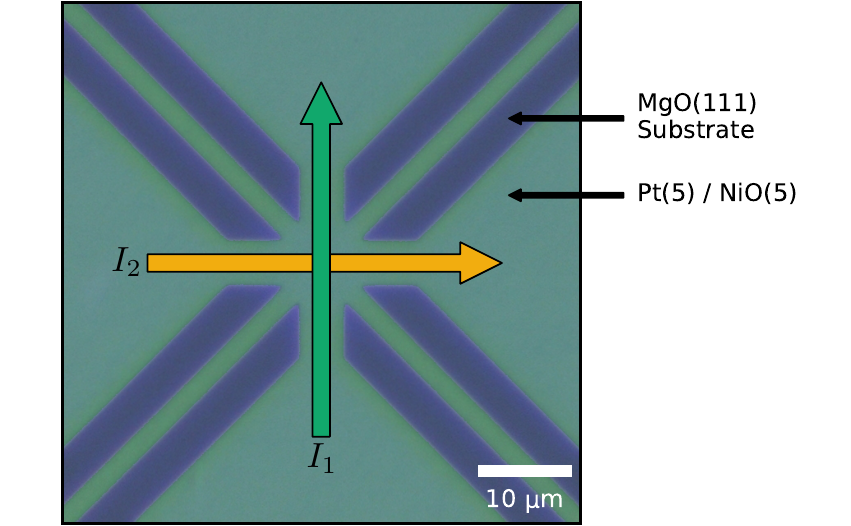}
    \caption{
        Micrograph of the device used for the electrical switching experiments.
        It consists of \ce{Pt} and \ce{NiO}, each \SI{5}{\nano\metre} thick, on top of an \ce{MgO(111)} substrate.
        The \txtclrA{} and \txtclrB{} arrow indicate the current pulse directions ($I_1$ \& $I_2$); the two diagonal lines are used for the spin-Hall magnetoresistance measurements.
        Colors have been enhanced for clarity.
    }
    \label{fig:device}
\end{figure}
For these experiments, we fabricate eight-terminal devices as shown in \cref{fig:device} consisting of \ce{NiO} on top of a \ce{Pt} layer (\SI{5}{\nano\metre} each) on \ce{MgO(111)} substrates.
Both layers are grown using DC magnetron sputtering, at \SI{565}{\celsius} for the \ce{Pt} layer to ensure the crystallinity of the \ce{Pt} layer and at \SI{430}{\celsius} in a 10:1 \ce{Ar}:\ce{O}-mixture for the \ce{NiO} layer.
The quality of these layers is ensured using a variety of characterization techniques (Supplemental Material~\cite{Supplementary}).
From these layers, devices as shown in \cref{fig:device} are fabricated using a combination of electron-beam lithography, electron-beam evaporation, lift-off, and ion-beam milling.

With these devices electrical switching experiments are performed, similar to the ones discussed in literature~\cite{Chen2018b}.
To switch the AF state, current pulses of \SI{3}{\milli\second} are applied along the wide orthogonal current lines $I_1$ and $I_2$, as indicated in \cref{fig:device} (the probe lines are also included when applying the pulses for a more homogeneous current distribution).
After every pulse, the small diagonal current lines are used for probing the present state with a Hall measurement; an alternating probing current of \SI{0.232}{\milli\ampere\rms} (equivalent to a current density of \SI{\sim 0.0186}{\ampere\rms\per\square\micro\metre}) at \SI{79}{\hertz} is passed in one direction and the generated Hall voltage is detected along the perpendicular current line using standard lock-in techniques.
Typically, there is a delay of \SI{5}{\second} between a pulse and the subsequent probing.

These experiments are performed within a cryostat, enabling the control of the temperature, that is placed in the center of a superconducting magnet, allowing for fields up to \SI{16}{\tesla} to be applied in the plane of the sample, along one of the probing lines (i.e.\ along the directions that are \SI{45}{\degree} rotated from the $\left<11\overline{2}\right>$ direction, approximately the $\left<\overline{0.7}\;2.7\;\overline{2}\right>$ and $\left<2.7\;\overline{0.7}\;\overline{2}\right>$ directions).

%   _____    ______    _____   _    _   _        _______    _____
%  |  __ \  |  ____|  / ____| | |  | | | |      |__   __|  / ____|
%  | |__) | | |__    | (___   | |  | | | |         | |    | (___
%  |  _  /  |  __|    \___ \  | |  | | | |         | |     \___ \
%  | | \ \  | |____   ____) | | |__| | | |____     | |     ____) |
%  |_|  \_\ |______| |_____/   \____/  |______|    |_|    |_____/

% \section{Results}
\begin{figure}
    \centering
    \includegraphics{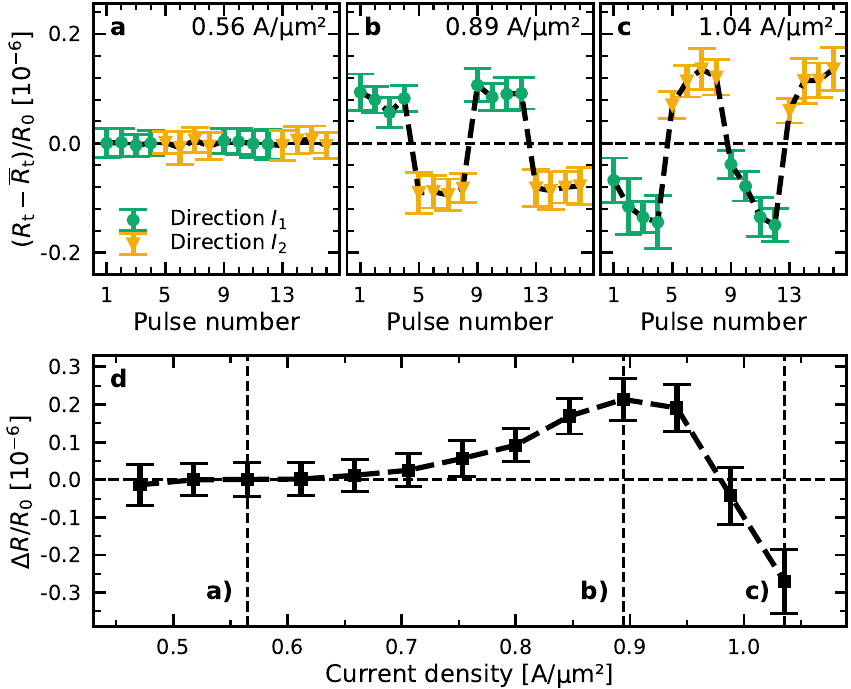}
    \caption{
        Current dependence of electrical switching in zero field at \SI{250}{\kelvin}.
        a) - c) show the measured transverse resistances (minus the average transverse resistance $\overline{R}_\mathrm{t}$) after a current pulse for increasing pulse current densities (\SI{0.56}{\currdens}, \SI{0.89}{\currdens}, and \SI{1.04}{\currdens}, respectively); the \txtclrA{} (\txtclrB{}) data points represent current pulses along direction $I_1$ ($I_2$).
        The shown data points are averages of 4 repeats of the full measurement cycle.
        d) The current dependence of the switching amplitude (i.e.\ the difference between the average transverse resistance of the two orthogonal current pulse orientations) as a function of the pulse current density.
        All data is normalized by the longitudinal device resistance $R_0=\SI{1.76(4)}{\kilo\ohm}$ and the black dashed lines are guides to the eye.
    }
    \label{fig:amplitude}
\end{figure}

A typical experimental result of electrical switching in the absence of a magnetic field is shown in \cref{fig:amplitude}a-c.
In the experiment, four current pulses are applied in one direction ($I_1$), whereafter four pulses are applied along the perpendicular direction ($I_2$).
After each pulse, the Hall resistance $R_\mathrm{t}$ along the small diagonal current lines is measured; for better comparison between different devices and temperatures, we subtract the average Hall resistance $\overline{R}_\mathrm{t}$ (averaged over the entire measurement) from the Hall resistance $R_\mathrm{t}$ and normalize the value to the longitudinal resistance $R_0$ (typically around \SI{1.7}{\kilo\ohm}, but the exact value can differ between devices, measurements and temperatures).
Changes of the Néel vector should show up in the Hall resistance since one of its contributions, the transverse component of the spin-Hall magnetoresistance (SMR), is sensitive to reorientations of the Néel vector of \ce{NiO} from \SI[retain-explicit-plus]{+45}{\degree} to \SI{-45}{\degree} with respect to the direction of the probing current~\cite{Chen2016}.
It is observed that for a specific pulse current density (\SI{0.89}{\currdens}) the two different pulse directions result in two distinct values of the Hall resistance, which is expected if indeed the Néel vector of \ce{NiO} has been switched by \SI{90}{\degree} by the current pulse.
However, it should be noted that, as mentioned earlier, other, non-magnetic effects can give similar results.

Upon increasing the pulse amplitude, as shown in \cref{fig:amplitude}c, we find that the behavior changes from a step-like switching at \SI{0.89}{\currdens} to a sawtooth-like switching at \SI{1.04}{\currdens}, where each subsequent current pulse in a certain direction still contributes to the final resistance state.
This sawtooth-like switching behavior also has an inverted sign when compared to the step-like switching at lower current densities; for lower current density the pulses marked in \txtclrA{} result in a high relative resistance state and the \txtclrB{} in a lower state, whereas for higher current density the \txtclrA{} pulses result in a low resistance state and the \txtclrB{} in a higher state.
For lower current densities (see \cref{fig:amplitude}a for \SI{0.56}{\currdens}) the resistance state is insensitive to the direction of the pulses; for this current density, no switching behavior is observed.

The total current density-dependent switching behavior is summarized in \cref{fig:amplitude}d, where the switching amplitude $\Delta R$ is plotted.
This switching amplitude is defined as the difference between the (average) transverse resistance after a current pulse in either direction, i.e.\ $\Delta R = \overline{R}_\mathrm{t,\,I_1} - \overline{R}_\mathrm{t,\,I_2}$; for consistency between devices, all values are normalized by the longitudinal resistance of the device $R_0$.
Here it becomes clear that starting from \SI{0.5}{\currdens} a form of current-induced switching becomes visible, reaching a maximum at around \SI{0.9}{\currdens}.
When the current density further increases, the switching amplitude starts to rapidly decrease, changes sign and increase again (in the negative direction).

This switching behavior, with the change of sign and the associated transition from step-like to sawtooth-like switching, indicates that there are at least two mechanisms at play in these experiments.
As mentioned earlier, these mechanisms can be either magnetic or non-magnetic of origin~\cite{Chiang2019,Churikova2020,Matalla-Wagner2020}.
However, from this set of measurements alone it is impossible to tell if one of the mechanisms involved in the experiment is or is not magnetic of origin, and if so, which part of the curve is explained by this magnetic mechanism.

To try to separate the magnetic from the non-magnetic effects, we perform an identical experiment in a superconducting magnet where magnetic fields up to \SI{16}{\tesla} can be reached.
When a sufficiently strong magnetic field is applied to the sample, the Néel vector (or equivalently, the individual magnetic moments) of the \ce{NiO} layer will be forced in a configuration perpendicular to the magnetic field~\cite{Machado2017,Fischer2018}.
Note that this resembles the state after a spin-flop transition; however, thin films of \ce{NiO} are known to rather undergo a monodomainization transition due to the relaxation of stress~\cite{Fischer2018,Gray2019}.
Upon applying a current pulse in such a magnetic field, we conjecture that one of two things will happen.
One possibility is that the current pulse will not be sufficiently strong to overcome the external magnetic field and influence the orientation of the Néel vector.
Alternatively, if the current pulse can affect the orientation of the Néel vector despite the presence of a strong external magnetic field, we expect the new orientation of the Néel vector will be reverted to the field-dominated orientation after the current pulse has stopped since the magnetic field is strong enough to force an orthogonal~\footnote{Although the high magnetic fields that are used in these experiments are expected to cant the magnetic moments slightly away from a direction that is perfectly orthogonal to the external magnetic field, we estimate that this canting is sufficiently small (\SI{\sim 0.4}{\degree} at \SI{15}{\tesla}) that this deviation can be disregarded in the interpretation of the experiments.} orientation of the magnetic moments~\cite{Machado2017,Fischer2018}.

\begin{figure}
    \centering
    \includegraphics{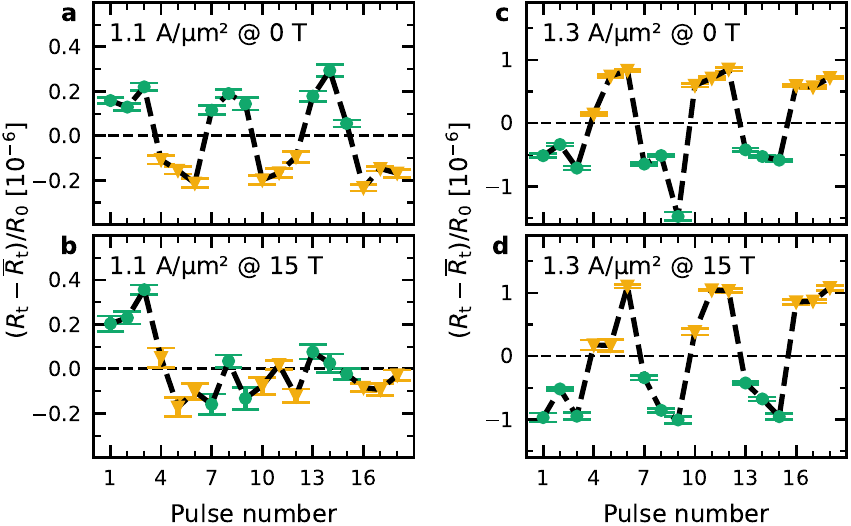}
    \caption{
        Switching of \ce{NiO} for multiple magnetic fields (top: \SI{0}{\tesla}, bottom: \SI{15}{\tesla}) and current pulse amplitudes (left: \SI{1.1}{\currdens}, right: \SI{1.3}{\currdens}) at \SI{240}{\kelvin}.
        The \txtclrA{} (\txtclrB{}) data-points represent the resistance-states (minus the average transverse resistance $\overline{R}_\mathrm{t}$) after pulses in direction $I_1$ ($I_2$).
        All data is normalized by the longitudinal device resistance $R_0=\SI{1.74(1)}{\kilo\ohm}$ and the black dashed lines are guides to the eye.
    }
    \label{fig:field}
\end{figure}

The results of these measurements are shown in \cref{fig:field}, both for a medium current density (a and b) and for a higher current density (c and d).
Note that these measurements have been performed at a different temperature (\SI{240}{\kelvin} rather than \SI{250}{\kelvin}).
Since both the magnetic and non-magnetic effects are dependent on temperature~\cite{Grzybowski2022}, the exact current densities that are required to obtain the same switching patterns can differ between different temperatures.
For this reduced temperature, a slightly higher current density is needed to obtain the same switching patterns as in \cref{fig:amplitude}; this is more thoroughly explored in the Supplemental Information~\cite{Supplementary}.

Similar to the previous measurements (\cref{fig:amplitude}) we can see that the current pulses along the different directions ($I_1$ and $I_2$) result in a switching pattern.
Please note that the high magnetic field setup results in a higher noise level in the measurements leading to a bit more irregular patterns than in zero field; this has no impact on the conclusions of this work as these are based only on the general difference between the two current directions.
When comparing the zero field experiment (\cref{fig:field}a) to the high field experiment (\SI{15}{\tesla}, \cref{fig:field}b) for the medium current density, we observe that the switching behavior that is visible at zero field is absent for the high field experiment.
For the high current density (\cref{fig:field} c and d, respectively) no such trend is observed; for both zero and high magnetic fields, switching behavior is observed.

As the electrical switching of \ce{NiO} is expected to be suppressed by a high magnetic field, we conjecture that the behavior we observe at medium current density is indeed caused by a magnetic effect; the switching behavior at higher current densities is then caused by non-magnetic effects as these are not expected to be affected by the magnetic field.
Note, however, that these two effects are not mutually exclusive and may have a gradual transition from one to the other.
Hence, it is possible that in the high current density regime there is a small magnetic component, and that in the medium current density regime there is a small non-magnetic component.

\begin{figure}
    \centering
    \includegraphics{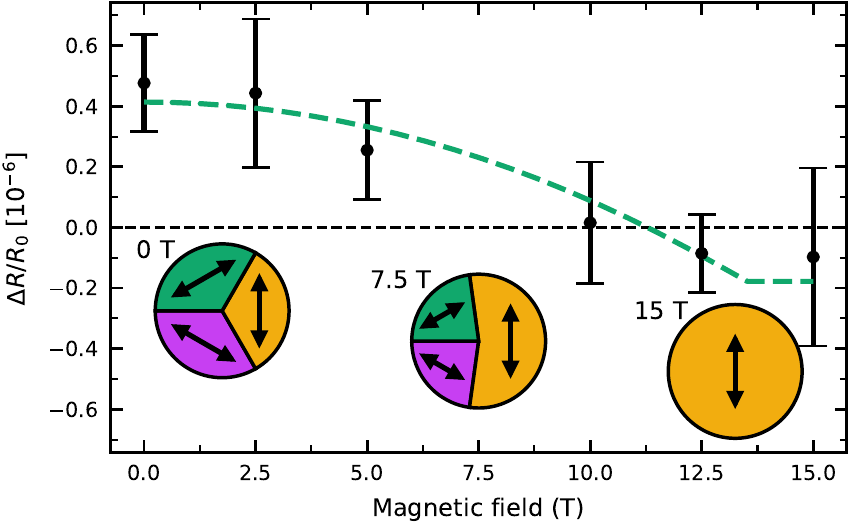}
    \caption{
        Evolution of the switching amplitude as a function of the external magnetic fields at \SI{240}{\kelvin}.
        The data points represent the measured switched states, averaged over every first pulses in that direction; the dashed \txtclrA{} line represent the fitted magnetic field-dependence \cref{eq:mdmodel}.
        Due to the increased noise level in the high magnetic field setup, the standard deviation (error bars) for these measurements are relatively large; nevertheless, a clear field-dependent trend is visible.
        The pie-charts represent the modelled distribution of the AF domains for \SI{0}{\tesla}, \SI{7.5}{\tesla}, and \SI{15}{\tesla}, from left to right; the arrows indicate the orientation of the magnetic moments in that domain.
    }
    \label{fig:model}
\end{figure}

To better identify these small contributions, we plot the switching amplitudes as a function of the external magnetic field magnitude in \cref{fig:model} for the medium current density.
In this figure, the dependence of the switching amplitude on the external magnetic field is visible.
At low magnetic fields, the switching amplitude is nearly unperturbed by the external magnetic field.
However, towards higher magnetic fields the switching amplitude is greatly reduced until it reaches a plateau around \SI{\sim 12}{\tesla}.

We conjecture that this behavior demonstrates the presence of two entangled contributions, the magnetic one that gets suppressed by the high field and the non-magnetic one that can be associated with the plateau that is reached at high fields.
To substantiate this hypothesis, modelling of the population of the antiferromagnetic domains for \ce{NiO} in high magnetic fields, and the resulting SMR is needed; for this, we make use of the monodomainization model~\cite{Fischer2018}.

The model describes the \ce{NiO(111)} layer of our samples by separating it into three domains, where the Néel vectors of the three domains are separated by \SI{120}{\degree} (corresponding to the three spin domains in the \ce{NiO(111)} plane~\cite{Uchida1967}); in \cref{fig:model} this distribution is represented in pie-charts.
While this interpretation and the analytical expressions that are derived from the model are based on \ce{NiO(111)} having three domains, the results are also valid in cases where the system has more (or equivalently, less well-defined) domains (Supplemental Material~\cite{Supplementary}), as is expected for the thin film of \ce{NiO} in our samples~\cite{Gray2019}.

In low magnetic fields, due to de-stressing, each of the domains takes up approximately an equal portion of the sample.
When a strong magnetic field is applied, the distribution of these domains changes due to the Zeeman energy, where the domain with (the largest projection of) the Néel vector perpendicular to the external magnetic field are preferred; this domain will increase in size at the cost of the other domains.
Finally, if the magnetic field surpasses the monodomainization field $B_\mathrm{md}$, there is only a single domain left, namely the domain with the Néel vector perpendicular to the external magnetic field.

This behavior results in a field-dependent transverse spin-Hall magnetoresistance (SMR) $R_\mathrm{SMR}$ that is given by~\cite{Fischer2018}:
\begin{equation}
    \label{eq:smr}
    R_\mathrm{SMR} \propto 
    \left\{
        \begin{array}{ll}
            \frac{B^2}{B_\mathrm{md}^2} & \mbox{if } B < B_\mathrm{md} \\
            1 & \mbox{if } B \geq B_\mathrm{md}
        \end{array}
    \right.
    .
\end{equation}

In the electrical switching experiments, it is expected that only a small portion of the magnetic structure is actually switched between the available domains upon applying a current pulse~\cite{Gray2019}.
However, when an external magnetic field is applied and the distribution of domains changes, fewer domains are available to switch between.
Therefore, we assume that the magnetic contribution to the switching signal scales as $R_\mathrm{SMR, max} - R_\mathrm{SMR}$.
An additional field-independent offset $R_\mathrm{nm}$ has been added to account for non-magnetic effects that are expected to contribute to the measured switching amplitude.
When combined with \cref{eq:smr}, the switching amplitude can then be expressed as:
\begin{equation}
    \label{eq:mdmodel}
    \Delta R = 
    \left\{
        \begin{array}{ll}
            R_\mathrm{nm} - R_\mathrm{m} \left(1 -\frac{B^2}{B_\mathrm{md}^2}\right) & \mbox{if } B < B_\mathrm{md} \\
            R_\mathrm{nm} & \mbox{if } B \geq B_\mathrm{md}
        \end{array}
    \right.
    ,
\end{equation}
where $R_\mathrm{m}$ is the SMR proportionality constant that accounts for the contribution of the changes in the antiferromagnetic state to the measured switching amplitude.

To fit the model to the data, the monodomainization field $B_\mathrm{md} = \SI{13.5}{\tesla}$ \cite{Supplementary} is kept constant and only $R_\mathrm{m}$ and $R_\mathrm{nm}$ are varied.
As shown in \cref{fig:model}, the fitted curve follows the data closely and emphasizes the gradual suppression of the magnetic contribution up to the monodomainization field and the non-magnetic plateau for higher fields.

As we now understand the field dependence of the switching amplitude $\Delta R$, it can be used to help settle the debate regarding the contributions of the magnetic and non-magnetic, parasitic effects to the observed switching signal.
To directly connect to the earlier discussed current dependence (\cref{fig:amplitude}), we repeat the switching in high magnetic field experiments for a series of current densities.
By fitting the monodomainization model (\cref{eq:mdmodel}) to the measurements at every current density, a measure for both the magnetic ($R_\mathrm{m}$) and the non-magnetic ($R_\mathrm{nm}$) contribution to the signal at zero magnetic field can be obtained.

\begin{figure}
    \centering
    \includegraphics{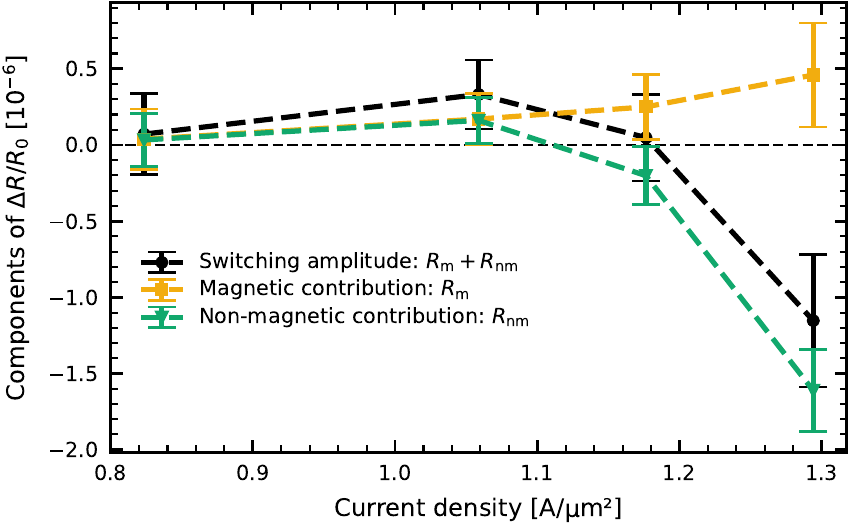}
    \caption{
        Separated magnetic (\txtclrB{}) and non-magnetic (\txtclrA{}) contributions to the electrical switching experiments as a function of pulse current density at \SI{240}{\kelvin}.
        The two contributions have been extracted from the magnetic field-dependence of the switching amplitude for each current density by fitting the data with \cref{eq:mdmodel} for $R_\mathrm{m}$ and $R_\mathrm{nm}$.
        The full switching amplitude (black circular marks) is the sum of $R_\mathrm{m}$ and $R_\mathrm{nm}$ and can be compared to the switching amplitude in \cref{fig:amplitude}d.
    }
    \label{fig:field_current}
\end{figure}

In \cref{fig:field_current} these separate contributions are plotted as a function of the current density of the pulses.
Here, it can be noted that the magnetic component slowly increases with increasing current density.
Simultaneously it is observed that the non-magnetic part rises (with an opposite sign from the magnetic part) from \SI{0.8}{\currdens} to \SI{\sim 1.05}{\currdens}; for higher current densities, however, it changes sign and the magnitude rises strongly in a non-linear manner, which hints at a thermal origin of this effect~\cite{Grzybowski2022}.

The sum of the two contributions shows (qualitatively) the same behavior as the switching amplitude at zero magnetic field (\cref{fig:amplitude}d), despite a small mismatch in temperature (\SI{250}{\kelvin} for \cref{fig:amplitude}d compared to \SI{240}{\kelvin} for \cref{fig:field_current}, related to different temperature control possibilities in the different setups).
For medium current densities, this indicates that the observed (step-like) switching signal (e.g.\ \cref{fig:amplitude}b) is a combination of both magnetic and non-magnetic components.
At high current densities (e.g.\ \cref{fig:amplitude}c), the (sawtooth-like) switching signal is mostly a result of the non-magnetic contributions; the inverted sign of this switching signal is also a consequence of these non-magnetic effects.
These conclusions are in line with earlier interpretations~\cite{Baldrati2019,Cheng2020} that the step-like switching can often be attributed to magnetic effects and sawtooth-like switching to non-magnetic effects.

We are convinced---in particular, based on the agreement with the expected behavior and the high magnetic field, typical for antiferromagnetic \ce{NiO}, needed to suppress the switching---that these results show that we are indeed able to separate the magnetic and non-magnetic effects using high magnetic fields.
Additional control experiments could be performed to further support these conclusions; to enable a meaningful comparison with the present data, a careful detailed analysis of results from non-magnetic control samples would be required.

%    _____    ____    _   _    _____   _        _    _    _____   _____    ____    _   _
%   / ____|  / __ \  | \ | |  / ____| | |      | |  | |  / ____| |_   _|  / __ \  | \ | |
%  | |      | |  | | |  \| | | |      | |      | |  | | | (___     | |   | |  | | |  \| |
%  | |      | |  | | | . ` | | |      | |      | |  | |  \___ \    | |   | |  | | | . ` |
%  | |____  | |__| | | |\  | | |____  | |____  | |__| |  ____) |  _| |_  | |__| | | |\  |
%   \_____|  \____/  |_| \_|  \_____| |______|  \____/  |_____/  |_____|  \____/  |_| \_|

% \section{Conclusion}
In summary, we have investigated the field-dependence of the electrical switching of \ce{NiO} to distinguish between the magnetic and non-magnetic contributions to the electrical switching experiments.
We showed that for lower current densities the (step-like) switching pattern is greatly suppressed upon applying a high magnetic field.
For higher current densities, on the other hand, the (inverted, sawtooth-like) switching pattern remains largely unaffected by the magnetic field, hinting at a non-magnetic origin.

Using a monodomainization model the magnetic and non-magnetic contributions can be separated from each other.
This confirmed that the switching at lower current densities is in part due to actual magnetic switching of \ce{NiO}; however, also at these lower current densities, a non-magnetic contribution is present.
Moreover, it showed that the change-of-sign that is observed at higher current densities is caused by non-magnetic effects; for these higher current densities there is still a magnetic part (which is slightly stronger than the magnetic effects at lower current densities) but the non-magnetic component dominates the signal.

We have shown that a strong magnetic field can be used to quantitatively disentangle magnetic and non-magnetic effects in experiments aiming at the electrical switching experiments of antiferromagnets.
This technique relieves the necessity for imaging the antiferromagnetic structure as a way to study the electrical switching of antiferromagnets and thereby opens a way to investigate more complex devices where the antiferromagnetic layer is just one of the many layers and imaging of the antiferromagnetic domains becomes increasingly difficult or impossible.

\begin{acknowledgments}
    We acknowledge S.\ Peeters for measuring the sheet resistance of an unpatterned sample.
    Sample fabrication was performed using NanoLabNL facilities.
    The research performed here was funded by the Dutch Research Council (NWO) under grant No.\ 680-91-113.
    This work was supported by HFML-RU/NWO-I, member of the European Magnetic Field Laboratory (EMFL).
    C.F.S.\ and M.J.G.\ conceived the experiments, fabricated the samples with help from T.J.K., and performed the measurements while advised by H.J.M.S.
    High magnetic field experiments were performed by C.F.S.\ and M.J.G.\ with help from K.R.\ and M.E.B.\ while advised by U.Z.
    C.F.S.\ performed the data analysis while advised by M.J.G., H.J.M.S., and R.D.
\end{acknowledgments}

\bibliography{Bibliography,BibliographyMisc}

\end{document}

% TODO: check harcoded references to main article
\title{
    Supplemental material for
    \texorpdfstring{\\}{ }
    Disentangling electrical switching of antiferromagnetic NiO using high magnetic fields
}
\date{\today}

\newcommand{\TUe}{Department of Applied Physics, Eindhoven University of Technology, P.O.\ Box 513, 5600 MB Eindhoven, the Netherlands}
\newcommand{\HFML}{High Field Magnet Laboratory (HFML - EMFL), Radboud University, 6525 ED Nijmegen, The Netherlands}
\newcommand{\ITP}{Institute for Theoretical Physics, Utrecht University, Leuvenlaan 4, 3584 CE Utrecht, The Netherlands}

\newcommand{\IEP}{Institute of Experimental Physics, Faculty of Physics, University of Warsaw, ul.\ Pasteura 5, PL-02-093 Warsaw, Poland}
\newcommand{\MagLab}{National High Magnetic Field Laboratory, Los Alamos National Laboratory, Los Alamos, NM 87545, USA}

\newcommand{\txtclrA}{green}
\newcommand{\txtclrB}{orange}

\author{C.\,F.\ Schippers}
\email{c.f.schippers@tue.nl}
\affiliation{\TUe}

\author{M.\,J.\ Grzybowski}
\altaffiliation[Current address: ]{\IEP}
\affiliation{\TUe}

\author{K.\ Rubi}
\altaffiliation[Current address: ]{\MagLab}
\affiliation{\HFML}

\author{M.\,E.\ Bal}
\affiliation{\HFML}

\author{T.\,J.\ Kools}
\affiliation{\TUe}

\author{R.\,A.\ Duine}
\affiliation{\TUe}
\affiliation{\ITP}

\author{U.\ Zeitler}
\affiliation{\HFML}

\author{H.\,J.\,M.\ Swagten}
\affiliation{\TUe}

\maketitle
\section{Sample characterization}
The \ce{Pt} and \ce{NiO} layers that are used for the electrical switching experiments described in the main paper, are deposited using magnetron sputtering.
\ce{Pt} is deposited from a \ce{Pt} target in an \ce{Ar} environment; \ce{NiO} is deposited in a reactive manner from a pure \ce{Ni} target in a 10:1 \ce{Ar}:\ce{O2}-mixture.
To improve the crystallinity of the layers, the depositions are performed at elevated temperature, at \SI{565}{\celsius} for \ce{Pt} and \SI{430}{\celsius} for \ce{NiO}.

\begin{figure}
    \includegraphics{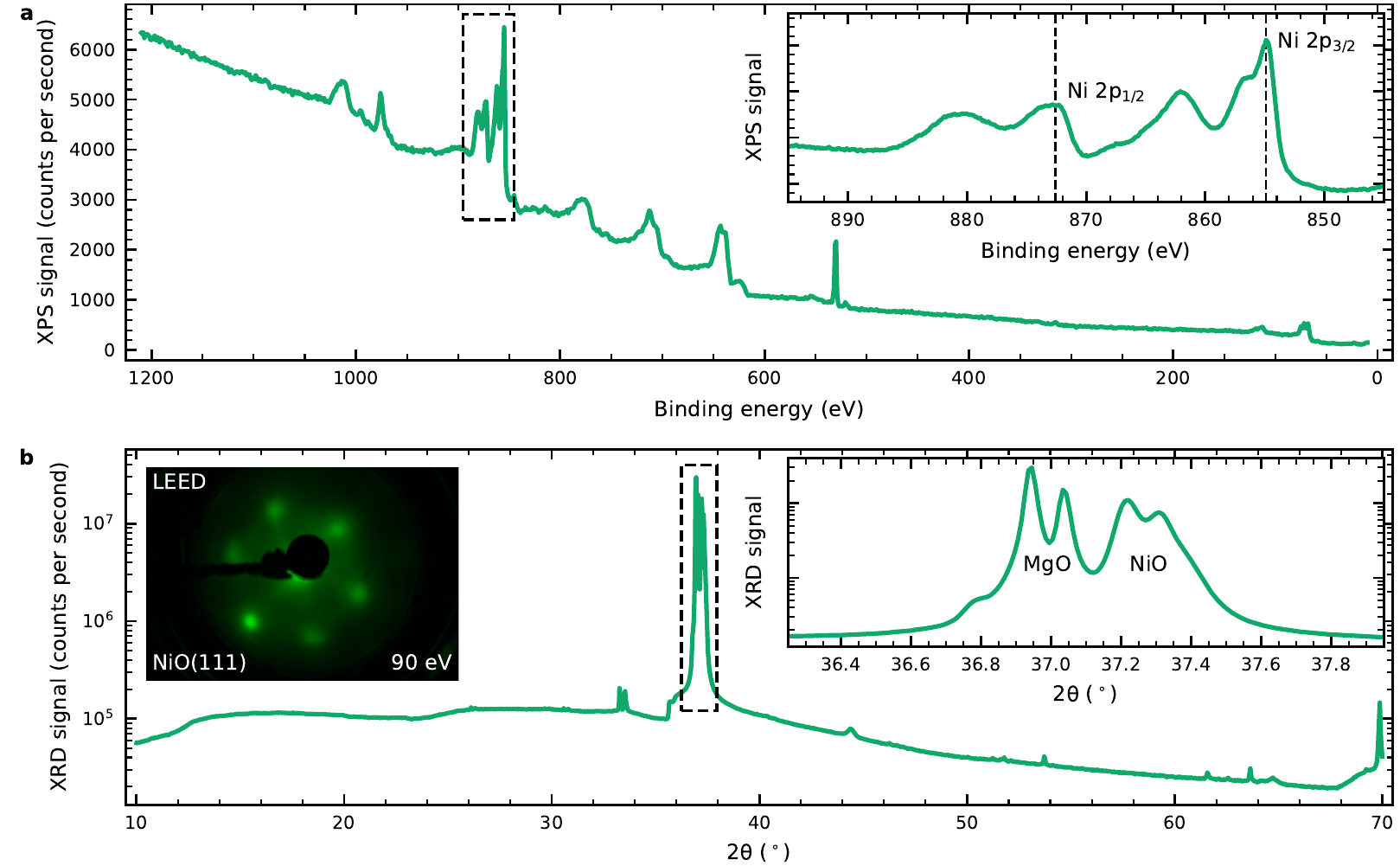}
    \caption{
        Characterization of \ce{NiO} as used for electrical switching experiments.
        (a) XPS scan of \ce{NiO} to investigate the stoichiometry, using an Al Kα source.
        The inset shows the region around the 2p-peaks of \ce{Ni}, showing the oxygen-induced satellites of the \ce{Ni} 2p peaks.
        (b) XRD scan of a \ce{MgO(111)/NiO(\SI{\sim500}{\nm})} sample, demonstrating the crystalline quality of the grown \ce{NiO} films; the inset shows the region around the (111) peaks of \ce{NiO} (around \SI{37.3}{\degree}) and \ce{MgO} (around \SI{37.0}{\degree}).
        For both \ce{NiO} and \ce{MgO}, the double peaks that are observed are a consequence of the x-ray source that provides x-rays of two wavelength (Cu Kα$_1$ and Cu Kα$_2$).
        The inset on the left shows the LEED pattern of an \ce{MgO(111)/Pt(5)/NiO(5)} sample at \SI{90}{eV}.
    }
    \label{fig:crystChar}
\end{figure}

These \ce{NiO} layers show a x-ray photoelectron spectroscopy (XPS) spectrum that is typical for NiO (\cref{fig:crystChar}a)~\cite{Mansour1994}.
The XRD spectrum of a \SI{\sim500}{\nm} thick \ce{NiO} layer (\cref{fig:crystChar}b), deposited using the same techniques, show \ce{NiO(111)} peaks, demonstrating the crystallinity of the grown \ce{NiO}.
In the thinner \ce{MgO/Pt(5)/NiO(5)} samples, the crystallinity is confirmed using low-energy electron diffraction (LEED, inset in \cref{fig:crystChar}b).

Using a four-point probe measurement~\cite{Smits1958}, we characterized the electrical properties of an \ce{MgO/Pt(5)/NiO(5)} sample, identical to the samples used for the electrical switching experiment that is described in the main paper.
The sample has a sheet resistance of \SI{45.4(5)}{\ohm\per\sq}; assuming that all current is shunted through the \SI{5}{\nm} thick \ce{Pt} layer (i.e.\ the resistivities of \ce{MgO} and \ce{NiO} are expected to be many times greater than that of \ce{Pt}), we find a resistivity of \ce{Pt} of \SI{22.7(3)}{\micro\ohm\cm}.
Although this values is significantly larger than the bulk resistivity of \ce{Pt} (\SI{10.8}{\micro\ohm\cm}~\cite{Arblaster2015}), the resistivity of thin \ce{Pt} films is known to be inversely dependent on the thickness of the film, where for \SI{5}{\nm} values of up to \SI{40}{\micro\ohm\cm} are observed~\cite{Boone2015,Nguyen2016}.

\begin{figure}
    \includegraphics{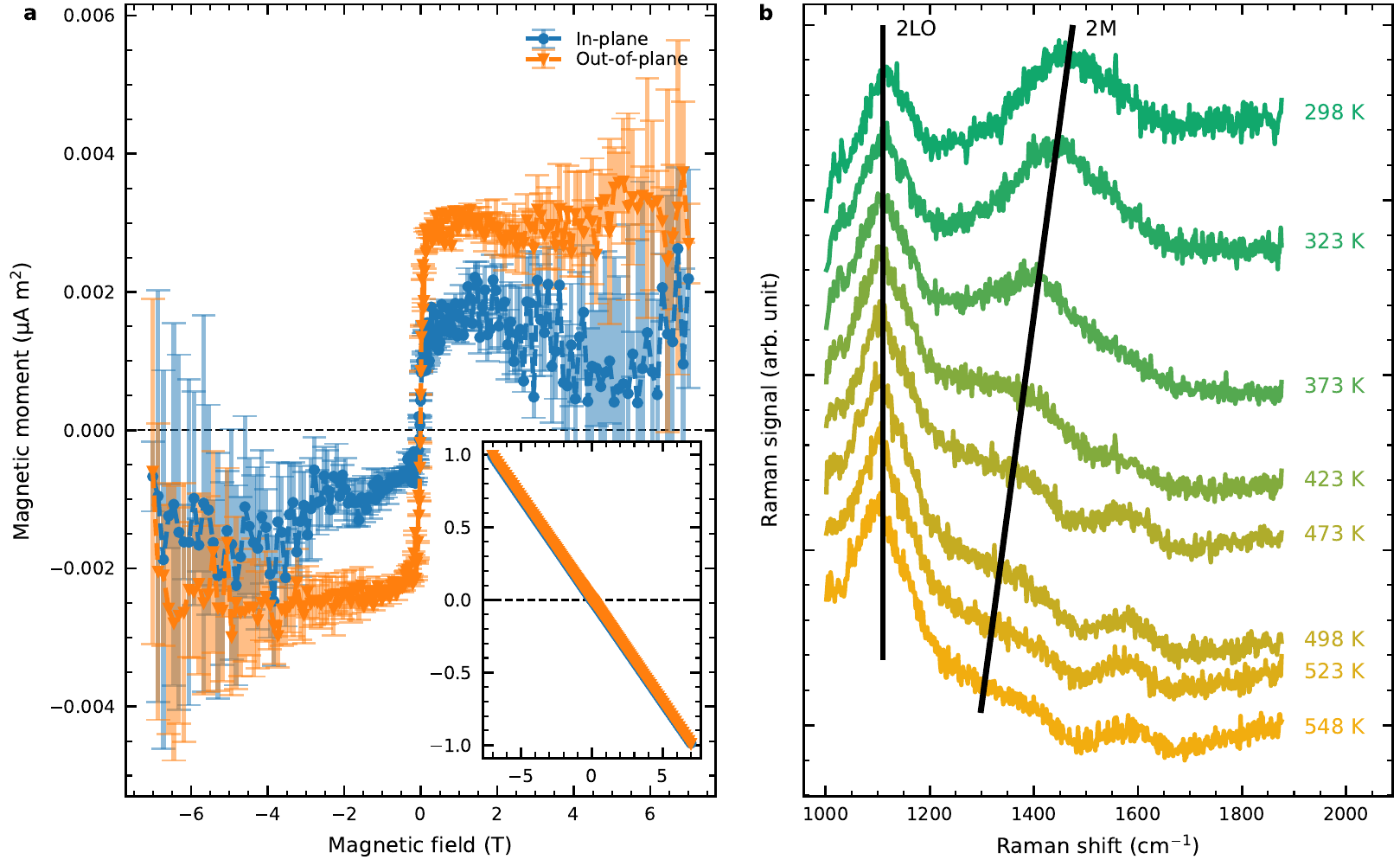}
    \caption{
        Magnetic characterization of \ce{NiO} as used for the electrical switching experiments.
        (a) Magnetic moment of an \ce{MgO/Pt(20)/NiO(60)} sample ($\SI{4.5}{\mm}\times\SI{4.5}{\mm}$) at \SI{300}{\kelvin} as a function of external magnetic field, measured using SQUID-VSM.
        A linear background (most likely originating from the diamagnetic response of the \ce{MgO} crystal) has been subtracted to show a small ferromagnetic contribution.
        The full measurement, with background, is shown in the inset; on this scale, the ferromagnetic contribution is not visible.
        (b) Raman spectroscopy measurement of an \ce{MgO/NiO(100)} sample for various temperatures.
        Around room temperature, two characteristic peaks are visible, one associated with the excitation or annihilation of two longitudinal optical phonons (indicated by `2LO'), and the other associated with the excitation or annihilation of two magnons (indicated by `2M').
        As emphasized by the black lines, for higher temperatures, the 2LO mode remains unchanged, but the 2M mode gets shifted and vanishes near the Néel temperature of \ce{NiO}.
    }
    \label{fig:magChar}
\end{figure}

Using superconducting quantum interferometer device vibrating sample magnetometer (SQUID-VSM), the magnetic moment of the sample (at \SI{300}{\kelvin} is measured for fields between \SI{\pm7}{\tesla}, as shown in \cref{fig:magChar}a.
The sample mostly shows a linear dependence on the external field (see inset); when a linear background is removed, a small ferromagnetic contribution can be observed, which can be attributed to either contamination of the sample or the sample-holder during the measurement, or to a ferromagnetic component present in the sample itself.
If this ferromagnetic component is a consequence of ferromagnetic \ce{Ni} ions (e.g. in clusters of unoxidised \ce{Ni}), it can be estimated that it has an equivalent volume of \SI{\sim 4e-06}{\mm\cubed} and would comprise only approximately \SI{0.3}{\percent} of the total \ce{NiO} volume (or one in every 180 \ce{Ni} ions behaves ferromagnetically).
Alternatively, it could be ascribed to the \ce{MgO} substrate, as \ce{Mg} or \ce{O} vacancies in \ce{MgO} can give rise to a small ferromagnetic moment~\cite{Khamkongkaeo2017}.

The actual antiferromagnetic structure of NiO is confirmed using Raman spectroscopy, \cref{fig:magChar}b.
At room temperature, the characteristic 2LO (two longitudinal optical phonon) and 2M (two magnon) modes~\cite{Lacerda2017} are observed; here, the 2M mode arises due to the creation (or annihilation) of two antiferromagnetic magnons in \ce{NiO}.
When the temperature is increased, this 2M peak moves to lower Raman shifts and starts to decrease in intensity; the peak has fully vanished near the Néel temperature of \ce{NiO} (\SI{523}{\kelvin}), demonstrating that \ce{NiO} is indeed antiferromagnetic.

\section{Determining the monodomainization field}
\begin{figure}
    \centering
    \includegraphics{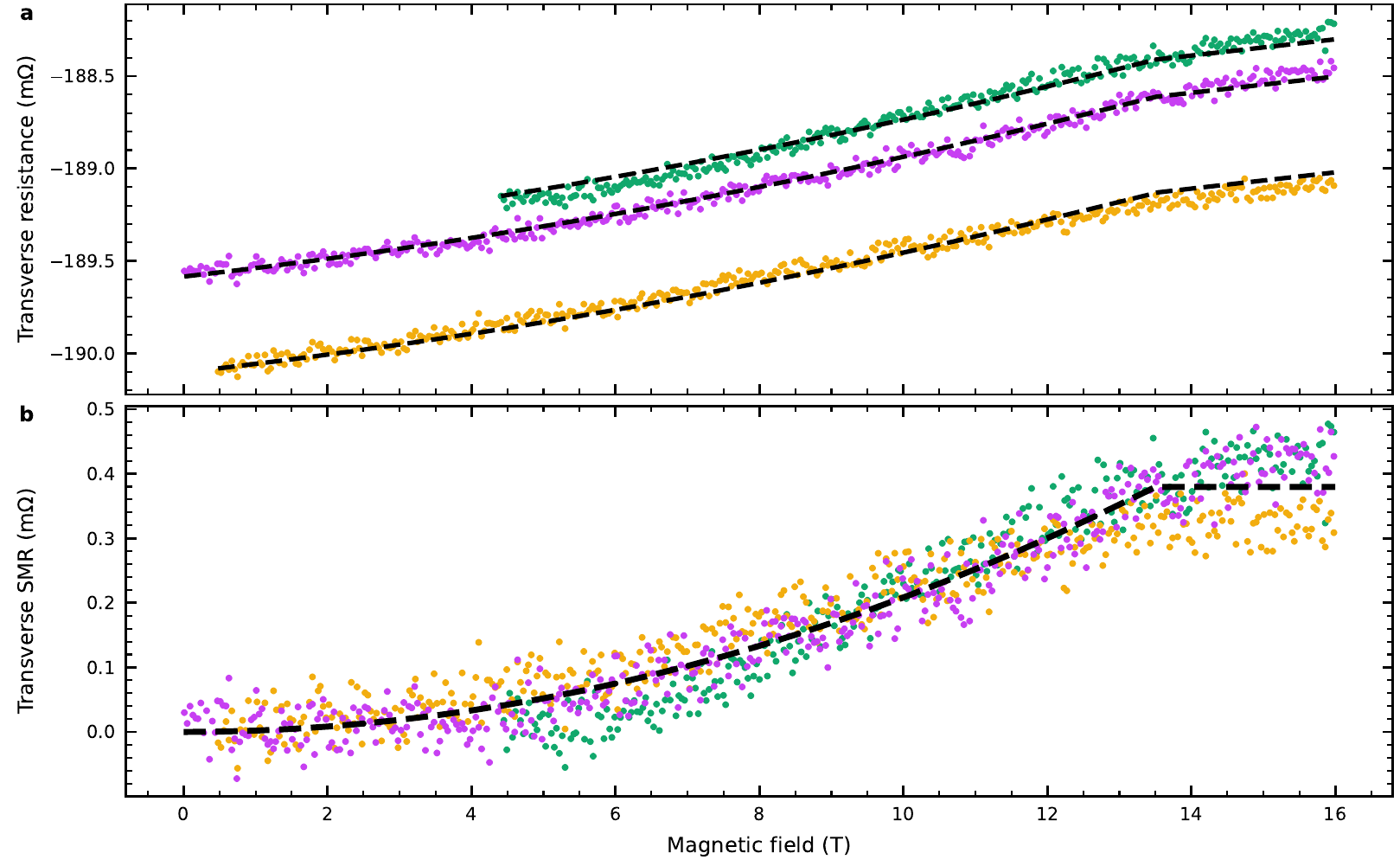}
    \caption{
        Transverse resistance measurement as a function of external magnetic field for determining the monodomainization field.
        Measurement was performed at \SI{240}{\kelvin}; the green and purple datasets are measured while increasing the magnetic field strength, the orange dataset while decreasing the magnetic field strength.
        Different coloured points represent multiple measurements on the same device.
        The data (a) was fitted with \cref{eq:smr_model} (black dashed lines), where $B_\mathrm{md}$, $R_1$, and $R_\mathrm{SMR}$ are shared for the different data sets and the offset $R_0$ is fitted separately for the different data sets.
        From this data, the pure SMR signal (b) is extracted by removing the background offset ($R_0$) and slope ($R_1$).
    }
    \label{fig:monodomainization}
\end{figure}

For fitting Eq.\ (2) of the main text to the data (Fig.\ 4 of the main text), the monodomainization field $B_\mathrm{md}$ is determined separately to reduce the number of fitting parameters.
As shown in \cref{fig:monodomainization}a, the transverse spin-Hall magnetoresistance is measured as a function of the external magnetic field~\cite{Fischer2018}.
From this the monodomainization field can be extracted by fitting the field-dependence of the transverse resistance $R_\mathrm{t}$, which includes the SMR, to the data:
\begin{equation}
    \label{eq:smr_model}
    R_\mathrm{t} = 
    \left\{
        \begin{array}{ll}
            R_0 + R_1 \frac{B}{B_\mathrm{md}} + R_\mathrm{SMR} \frac{B^2}{B_\mathrm{md}^2} & \mbox{if } B < B_\mathrm{md} \\
            R_0 + R_1 \frac{B}{B_\mathrm{md}} + R_\mathrm{SMR} & \mbox{if } B \geq B_\mathrm{md}
        \end{array}
    \right.
    .
\end{equation}
Here, $R_\mathrm{SMR}$ is the SMR contribution, and $R_0$ and $R_1$ account for a linear background that is caused by non-magnetic effects ($R_0$) and the ordinary Hall effect ($R_1$)~\footnote{Although the magnetic field is aligned in the plane of the sample, such that no contribution from the ordinary Hall effect is expected, a small misalignment of the sample can cause a small out-of-plane component of the magnetic field, and hence a small contribution of the ordinary Hall-effect to the transverse resistance.}.

The measurements shown in \cref{fig:monodomainization}a are fitted simultaneously for a more accurate value of $B_\mathrm{md}$; only the offset $R_0$ is different between the separate measurements, as the zero-field transverse resistance value is found to fluctuate between different measurements (possibly caused by variations the exact sample temperature between the measurements, since the \ce{Pt} resistivity is highly sensitive to the temperature).
Using the obtained values for the background contributions $R_0$ and $R_1$, the background can be removed to better see the field-dependent spin-Hall magnetoresistance in \cref{fig:monodomainization}b.
Here, we can see that the fitted curve matches closely with the separate measurements.
From this fit, we obtain a value of the monodomainization field of $B_\mathrm{md} = \SI{13.5(2)}{\tesla}$; although this value is in line with some values found in literature~\cite{Fischer2018}, the exact value is difficult to compare as it is known to depend on temperature, but also strongly on sample properties, such as layer thickness, substrate, adjacent layers, and general film quality~\cite{Fischer2018,Kimata2020}.

\section{Validity of the SMR model}
The analytical model used in the main text (Eqs.\ (1) and (2) of the main text) and in the previous section (\cref{eq:smr_model}) are based upon the assumption that the \ce{NiO(111)} can be divided into 3 antiferromagnetic domains, based on triaxial in-plane anisotropy of \ce{NiO(111)}~\cite{Fischer2018}.
Within this model, the Néel vectors of these three domains align, in the absence of an external field, such that there is an angle of \SI{120}{\degree} between them.
While this coincides with the three in-plane anisotropy axes, this is mostly a consequence of a de-stressing field coming from the coupling between the local Néel vectors and the crystal lattice; the anisotropy is expected to be much smaller than the de-stressing field.

However, for thin films it has been reported that the in-plane triaxial anisotropy vanishes, leading to a random (i.e.\ not adherent to any specific in-plane direction) orientation of the antiferromagnetic moments in the plane of the sample~\cite{Gray2019}.
To verify that the model that is used in the main text and the previous section remains valid for the thin films of \ce{NiO}, we compare the analytical model (derived within the assumption of 3 different domains) that we have used to a numerical simulation of the model for a system with 3 domains and a system with 120 domains to approximate the random distribution of domains.
For all calculations (including the derivation of the analytical model), it was assumed that the de-stressing field is significantly larger than the anisotropy field, such that the anisotropy field can be neglected~\cite{Fischer2018}.

\begin{figure}
    \centering
    \includegraphics{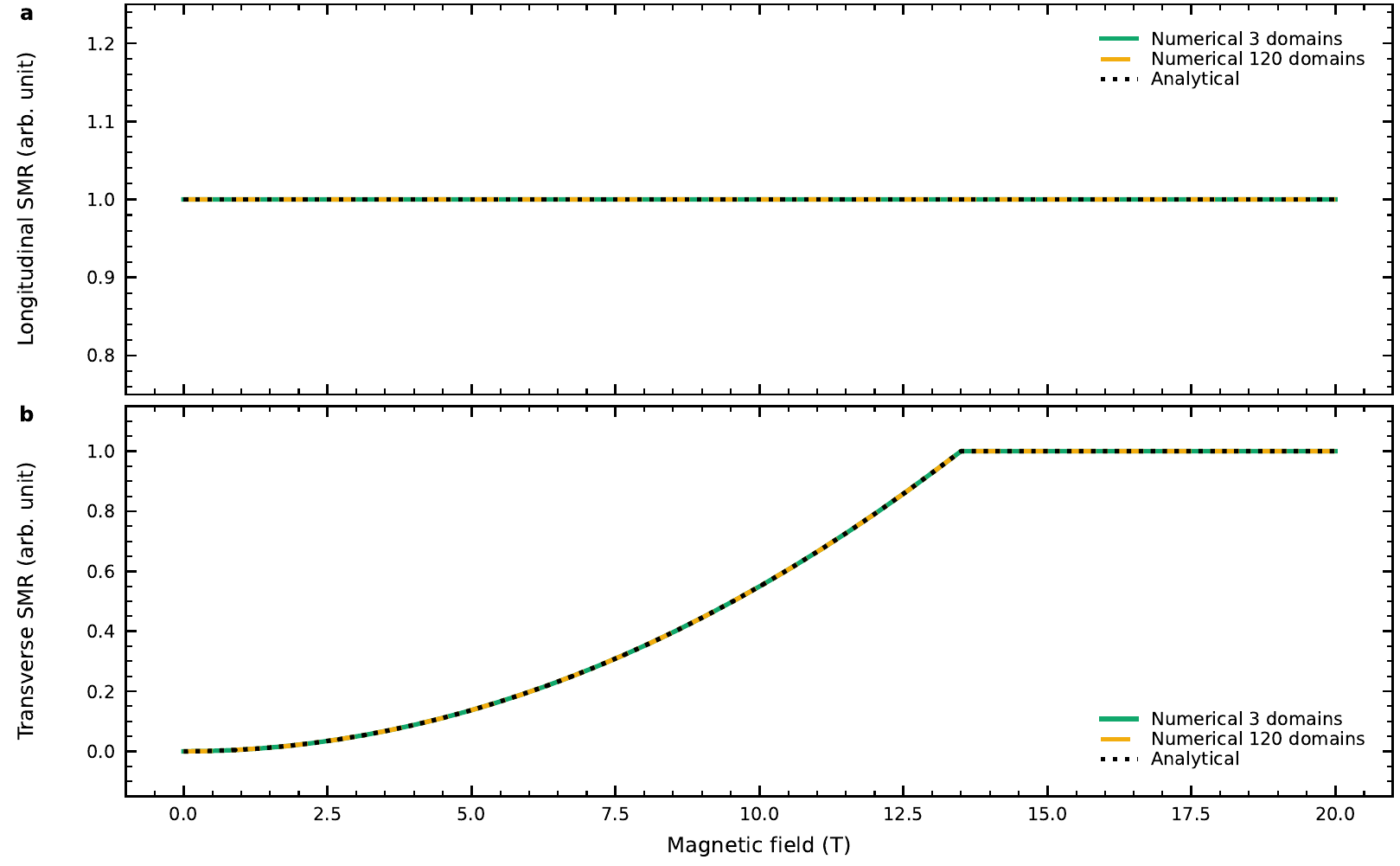}
    \caption{
        Comparison of the longitudinal (a) and transverse (b) SMR signal of a \ce{Pt/NiO} system as a function of magnetic field between numerical and analytical calculations.
        The \txtclrA{} (\txtclrB{}) lines are numerical calculations under the assumption the system in separated into 3 (120) domains.
        The black line is the analytical calculation, based on the model presented by \textcite{Fischer2018}, which is based upon a system that is separated into 3 domains.
    }
    \label{fig:smr_validity}
\end{figure}

\Cref{fig:smr_validity} shows the longitudinal and transverse SMR signal for the analytical model and for the numerical calculation with 3 and 120 domains.
Here we can see that both the longitudinal and transverse signals do not differ between the analytical model and both the numerical calculations.
The longitudinal SMR is for all cases insensitive to the external field (which is expected for the configuration of our experiments, i.e.\ when the external field is applied at an angle of \SI{45}{\degree} to the probing current).
For the transverse SMR, we can see that the quadratic behavior up to the monodomainization field $B_\mathrm{md}$ is also present for the three calculations.
This demonstrates that the analytical model can indeed also be used for thin films of \ce{NiO} where the in-plane triaxial anisotropy has vanished.

\section{Temperature dependence of electrical switching}
\begin{figure}
    \includegraphics{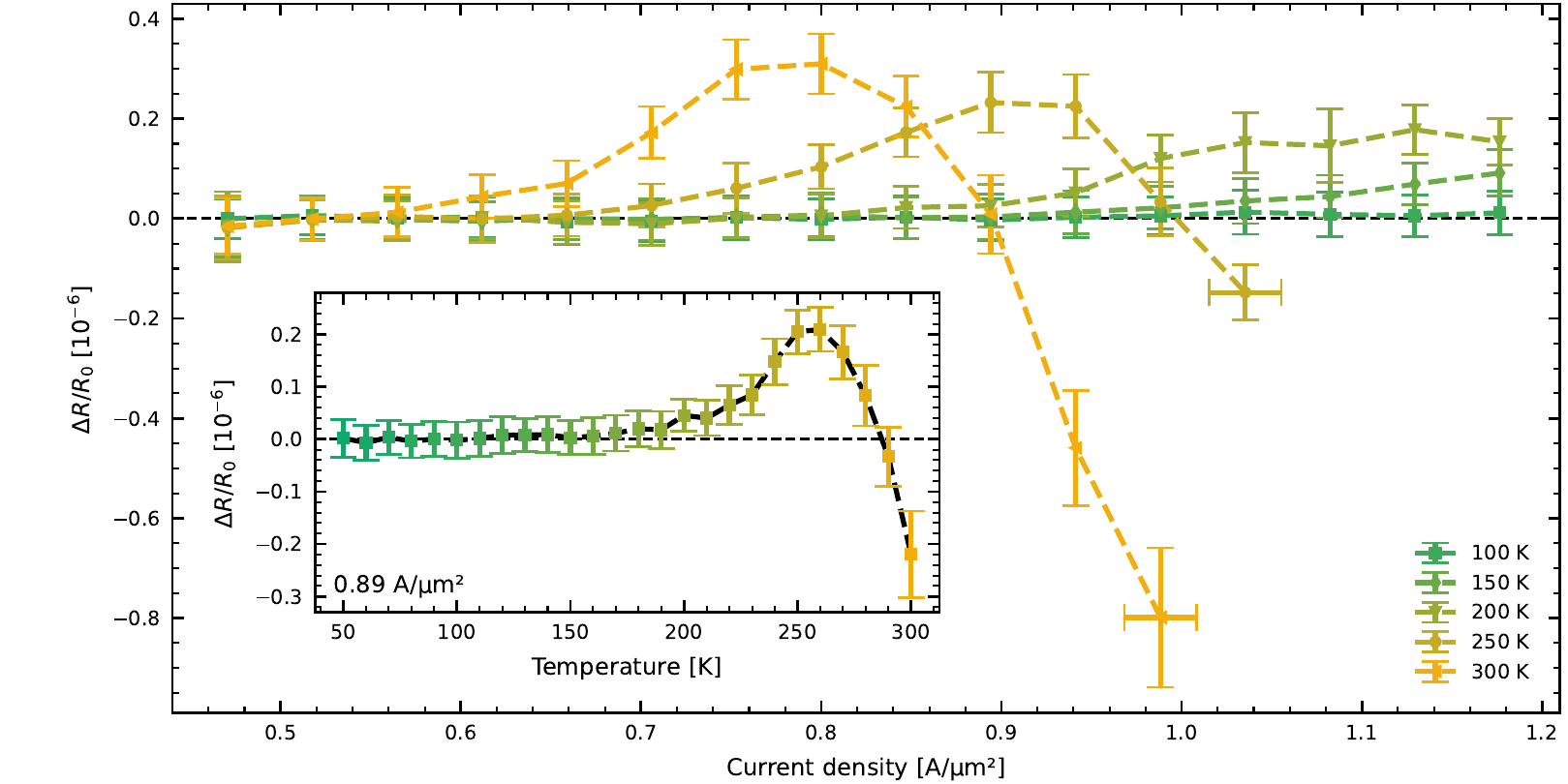}
    \caption{
        Switching amplitude (i.e.\ the difference between the average transverse resistance of the two orthogonal current pulse orientations, as defined in the main text) as a function of pulse current density, for various temperatures.
        The dashed lines are guides to the eye.
        For clarity, measurements for temperatures lower than \SI{100}{\kelvin} has been omitted since (within this range of current densities) these show the same behaviour as the curve for \SI{100}{\kelvin}.
        For \SIlist{250;300}{\kelvin}, the rightmost data point (i.e.\ highest current density), the voltage compliance is reached for the current pulse; for these temperatures, higher pulse current densities have been omitted.
        The inset shows the switching amplitude as a function of temperature for a fixed current density (\SI{0.89}{\currdens}); to emphasize the downward trend at higher temperatures, the switching amplitude is obtained using every third current pulse in a certain direction (as compared to every first current pulse for the other graphs).
    }
    \label{fig:tempSwitch}
\end{figure}

In the main text, the electrical switching experiments are performed at two temperatures, \SIlist{240;250}{\kelvin} (figs. 3 and 2 of the main text, respectively).
At these two temperatures, a different current density is required to obtain the same switching pattern: to obtain a positive, step-like switching pattern, at \SI{250}{\kelvin} (fig. 2b of the main text) a current density of \SI{0.86}{\currdens} is needed, while at \SI{240}{\kelvin} (fig. 3a of the main text) a current density of \SI{1.1}{\currdens} is used.

To investigate how the electrical switching is affected by the ambient temperature, \cref{fig:tempSwitch} shows the measured electrical switching amplitude (as defined in the main text) as function of the pulse current density for various temperatures.
Here, we note that the switching behavior that we observe in fig. 2d of the main text is again present for different temperatures, but presents itself at different current densities.
In the inset in \cref{fig:tempSwitch}, showing the switching amplitude as a function of ambient temperature for a fixed current density, this temperature dependence is clearly visual; where for a single current density, at one temperature a positive switching amplitude is observed, a negative switching amplitude can be observed when the temperature is increased.

This behavior can be understood by considering that the processes that are involved in the electrical switching either depend on (or are assisted by) thermal fluctuations (such as switching by spin--orbit torques, where an anisotropy energy barrier has to be overcome) or are thermal processes (such as thermomagnetoelastic switching or damage by joule heating).
Hence, these processes are easier and require lower pulse current densities at higher ambient temperatures.
This also explains the discrepancy between the current densities that are used for obtaining the same switching pattern between figs. 2 and 3 of the main text, since these experiments have been performed at different temperatures.

\bibliography{Bibliography}